\documentclass[12pt,dvips]{article}
\usepackage{graphicx}
\usepackage{latexsym}
\newcommand{\MSbar}{{\ensuremath{\overline{\textrm{MS}}}}}
\newcommand{\scriptsizeMSbar}{{\ensuremath{\overline{\textrm{\scriptsize MS}}}}}

\begin{document}
%%%%%%%%%%%%%%%%%%%%%%%%%%%%%%%%%%%%%%%%%%%%%%%%%%%%%%%%%%%%%%%%%%%%%%%%%%%

%%%%%%%%%%%%%%%%%%%%%%%%%%%%%%%%%%%%%%%%%%%%%%%%%%%%%%%%%%%%%%%%%%%%%%%%%%%
% TITLEPAGE
%%%%%%%%%%%%%%%%%%%%%%%%%%%%%%%%%%%%%%%%%%%%%%%%%%%%%%%%%%%%%%%%%%%%%%%%%%%
\thispagestyle{empty}
\date{\today}
\title{
%\vspace{-5.0cm}
%\begin{flushright}
%{\normalsize UNIGRAZ-}\\
%{\normalsize UTP-dd-dd-yy}\\
%\end{flushright}
\vspace*{0.05cm}
{\bf Renormalization of bilinear quark operators for the chirally 
improved lattice Dirac operator}}
\author
{Christof Gattringer$^1$, 
M. G\"ockeler$^{2,1}$, Philipp Huber$^3$ \\and C. B. Lang$^3$\\~\\
$^1$Institut f\"ur Theoretische Physik, \\
Universit\"at Regensburg, D-93040 Regensburg, Germany\\\\
$^2$Institut f\"ur Theoretische Physik, \\
Universit\"at Leipzig, D-04109 Leipzig, Germany\\\\
$^3$Institut f\"ur Theoretische Physik,
\\Universit\"at Graz, A-8010 Graz, Austria}
\maketitle
\begin{abstract}
We compute non-perturbative renormalization constants of fer\-mi\-o\-nic 
bilinears for the chirally improved lattice fermions in the quenched 
approximation of QCD. We address finite size effects and the influence of 
Gribov copies. Our results are presented in the
RI' and \MSbar~ schemes as well as in RGI form and we discuss 
relations between the renormalization constants implied by chiral symmetry. \\
After publication we corrected the numerator of the first coefficient of $\alpha_s^3$ in (\ref{renormalization_definition_Rm}) from 3696847 to 3890527, which yields a $0.2\%$ higher value of the conversion coefficient at $\mu=2\,\textrm{GeV}$.
\end{abstract}

% \vskip1cm

\newpage

\section{Motivation}
Lattice Dirac operators obeying the Ginsparg-Wilson constraint \cite{GiWi82}
(GWC) provide a lattice formulation of chiral symmetry closest to 
the continuum form; its violation is local and ${\cal O}(a)$. Exact 
GW-fermion fields are protected by their chiral 
symmetry \cite{Lu98} and renormalization of operators constructed
from them is simpler than for e.g.\ Wilson fermions. Chiral symmetry 
implies several relations between renormalization constants. Checking
these relations provides an important check of how well chiral
symmetry is implemented in a particular lattice formulation of QCD
\cite{othercalcs}.

To be more explicit, let us consider the local, flavor
non-singlet quark field bilinear operator
\begin{equation}
O_\Gamma \; \equiv \; {\overline u}\, \Gamma\, d \;. 
\label{eq_def_bilinear}
\end{equation}
Here $\Gamma$ denotes a Clifford algebra matrix. According to their Lorentz
symmetry\ we denote the five types of $\Gamma$ by S, V, A, T and P 
corresponding to scalar, vector, axial vector, tensor
and pseudoscalar 
(S$\sim\mbox{\boldmath{$\mathsf{1}$}}$, V$\sim\gamma_\mu$, 
A$\sim\gamma_\mu\gamma_5$,
T$\sim \frac{i}{2} [\gamma_\mu,\gamma_\nu]$, P$\sim\gamma_5$). 
Chiral symmetry implies
in particular $Z_S=Z_P$ and $Z_V=Z_A$. For conserved covariant
currents, Ward identities give $Z_V=1$.

So far only the overlap action \cite{NaNe93a,NaNe95} provides
exact GW-fermions. There are, however, several lattice Dirac operators
obeying the GWC approximately (such as fixed point \cite{HaNi94} and chirally 
improved (CI) fermions \cite{Ga01}) or in some limit 
\cite{Ka92,FuSh95}.
Although technically  more demanding than standard lattice fermion 
formulations they are still substantially less expensive (in terms of computer
resources) than exact overlap fermions. 
The Bern-Graz-Regensburg (BGR) collaboration \cite{GaGoHa04} has been studying
the low-lying hadron spectrum and other
properties for fixed point \cite{HaNi94} and CI fermions \cite{Ga01,GaHiLa00}.

Results involving renormalized quantities like the pion decay constant, the
chiral
condensate or the quark masses, need renormalization constants in order to be
connected to experiment, usually given in the framework of the \MSbar~(modified
Minimal Subtraction) renormalization scheme. Here we present and discuss 
results for the renormalization factors of bilinear operators of 
type (\ref{eq_def_bilinear}) for the CI 
Dirac operator. A subset of our results, but without discussion of the full
calculation, has been used earlier \cite{BrBuGa03}.

The chirally improved Dirac operator has been introduced in
\cite{Ga01,GaHiLa00,GaLa03} as an approximate solution of the Ginsparg-Wilson
constraint. It 
is formulated as a truncated series of interaction terms with
coefficients depending on the gauge coupling. Within the BGR collaboration it
has been used to obtain the ground state hadrons in the quenched approximation
\cite{GaGoHa04} down to pion masses of 270 MeV, and it
has been demonstrated to have good chiral properties \cite{GaGoHa03}. 
The renormalization constants for quark bilinears determined in this
work permit a more detailed analysis of the chiral properties of
the CI operator.

In Sect.\ \ref{sect_methods} we briefly 
discuss the method we use for calculating the
renormalization constants \cite{MaPiSa95} and its
implementation. We then summarize the technical details such as the 
parameters of the simulations, the chiral limit, volume dependence and
gauge fixing ambiguities in Sect.\ \ref{sect_technicalities}, 
before we present the lattice results
in Sect.\ \ref{sect_results}. In order to compare with continuum
notations like the \MSbar\ scheme and to discuss the renormalization
scale dependence, we recapitulate the conversion to 
other renormalization schemes 
in Sect.\ \ref{sect_conversions} and then discuss our results
for the converted renormalization constants. 
We conclude and summarize in Sect.\ \ref{sect_conclusion}.

\section{Method}\label{sect_methods}

We want to compute renormalization constants non-perturbatively on the lattice.
For this purpose we need a renormalization scheme which can be implemented 
in lattice Monte Carlo simulations as well as in continuum
perturbation theory. The latter property is necessary to enable the
conversion of the lattice results to a more conventional scheme 
such as \MSbar. A renormalization scheme satisfying these requirements is 
the so-called RI (regularization independent) scheme suggested by 
Martinelli et al.\ \cite{MaPiSa95}. 

In this scheme one studies expectation values of the bilinear quark
operators between quark fields at a specific momentum value $p^2=\mu^2$, 
\begin{equation}\label{def_RI}
\langle p\mid O_\Gamma\mid p\rangle \, \Big|_{p^2=\mu^2} 
\end{equation}
and matches them to the corresponding
tree-level matrix element $\langle p\mid O_\Gamma\mid p\rangle_0$. 
This procedure is expected to work in a window
\begin{equation}\label{mu_limits}
\Lambda_\textrm{\scriptsize QCD}^2 \ll \mu^2 \ll 1/a^2
\end{equation}
where discretization effects can be neglected, because the renormalization 
scale $\mu$ is small compared with the lattice cut-off $1/a$, 
and (few-loop) continuum perturbation theory can be used to connect 
different schemes, because $\mu$ is much larger than the QCD scale parameter
$\Lambda_\textrm{\scriptsize QCD}$. For comparing with the \MSbar\ scheme a 
typical value is $\mu=2\,\textrm{GeV}$. In
most calculations one has $a\,\mu\approx 1$ (or even somewhat larger) and the
upper limit is not strictly obeyed. On the other hand, the limit also
depends on the scaling properties of the actions involved.

Since (\ref{def_RI}) is gauge-variant, one has to work in a fixed gauge and
must compare the gauge dependent lattice matrix elements with the 
continuum results 
in the same gauge. Landau gauge fixing is a suitable choice, but one 
has to keep in mind that the Gribov copies uncertainty could spoil the 
comparison. In the lattice calculations one finds little, if any signal 
of such an effect  (see our discussion below in 
Sect.\ \ref{sect_technicalities} as well as 
\cite{PaPeTa94,GiPaPa01,GiPeTa02a}).

Let us briefly summarize the method following \cite{MaPiSa95} in the
modification of \cite{GoHoOe99}. When multiplying (\ref{def_RI})
with $\langle p\mid O_\Gamma\mid p\rangle_0^{-1}$ and taking the 
trace (note that the object in Eq.\ (\ref{def_RI}) is a matrix in 
color and Dirac space) one obtains for the renormalization condition
\begin{equation}
Z_\Gamma \; \frac{1}{12} \, tr [ \, \langle p\mid O_\Gamma\mid p\rangle \,
\langle p\mid O_\Gamma\mid p\rangle_0^{-1} \, ] \, \Big|_{p^2=\mu^2} 
\; = \; 1 \;.
\end{equation}
The matrix element 
\begin{equation}
\langle p\mid O_\Gamma\mid p\rangle \; = \; \frac{1}{Z_q} \, 
\Lambda_\Gamma(p)
\label{eq_matelement}
\end{equation}
is proportional to the amputated Green function 
\begin{equation}
  \Lambda_\Gamma(p) \; = \; 
S^{-1}(p) \; G_\Gamma(p) \; S^{-1}(p) \; ,
  \label{RI-MOM_vertex_function}
\end{equation}
and $Z_q$ is the quark field renormalization constant to be discussed below. 
%The ``dressed'' Green function
The Green function
$G_\Gamma(p)$ is determined as the expectation value
\begin{equation}      
  G_\Gamma (p)_{\alpha\beta} \; = \; 
\frac{1}{V}\, \sum_{x,y} \,e^{-i p (x-y)} \,
  \left\langle u_\alpha(x)\,
\sum_z \, O_\Gamma(z) \,\, \bar{d}_\beta(y) \right\rangle\;.
  \label{green_function}
\end{equation}
The indices $\alpha, \beta$ run over color and Dirac indices and $V$ 
denotes the lattice volume. The quark propagator is
\begin{equation}
  S_{\alpha\beta}(x,y) \; = \; 
\langle u_\alpha(x) \,\bar{u}_\beta(y) \rangle \; = \; 
\langle d_\alpha(x) \,\bar{d}_\beta(y) \rangle 
\end{equation}
(assuming that $u$ and $d$ have equal masses and using the Landau gauge 
for the expectation value). 

So we have to compute $G_\Gamma(p)$ and $S(p)$.
This is done in the following way. For the quark propagator $S_n$ evaluated
on a single gauge configuration $n$ we define
\begin{equation}
  S_n(x|p) \; = \; \sum_{y}\, e^{i p y}\, S_n(x,y) \; .
  \label{def_sn}
\end{equation}
Taking into account $\gamma_5$-hermiticity of the propagator we
may, for quark bilinear operators $O_\Gamma$ as defined in 
Eq.\ (\ref{eq_def_bilinear}), rewrite $G_\Gamma (p)$ in terms of the
quantities (\ref{def_sn}),
\begin{eqnarray}
  G_\Gamma (p) & = &\frac{1}{V}\, \sum_{x,y,z} \,e^{-i p (x-y)} 
 \, \langle u(x) \,\bar{u}(z) \,\Gamma\, d(z) \,\bar{d}(y) \rangle \\
  & \approx &\frac{1}{V\,N}\, \sum_{n=1}^N  
\sum_{z} \,\gamma_5 \,S_n(z|p)^\dagger \,\gamma_5 \,
       \Gamma\, S_n(z|p)\;,
  \label{RI-MOM_green_function}
\end{eqnarray}
averaging over $N$ gauge configurations. Similarly we find
the quark propagator in momentum space
\begin{equation}
  S(p) \; \approx\; \frac{1}{VN}\, 
\sum_{n=1}^N\, \sum_{x} e^{-i p x}\, S_n(x|p) \;.
\end{equation}
$S_n(y|p)$ is computed by 
solving the lattice Dirac equation ($D_{CI}$ denotes the chirally
improved Dirac operator)
\begin{equation}
  \sum_{y}\, D_{CI}(z,y) \,S_n(y|p) \; = \;  e^{i p z}
  \label{RI-MOM_lattice_dirac_equation}
\end{equation}
with a momentum source (cf. \cite{GoHoOe99}). This has the disadvantage that
one has to determine the quark propagators for several momentum sources,
whereas in the original method \cite{MaPiSa95} one uses point sources (i.e.\
taking into account just $z=0$ instead of summing over all $z$) and projects
the quark sink to the desired momentum values. However, using momentum
sources has the big advantage of a significantly better signal. 

The quark field renormalization constant is obtained by comparing
the quark propagator to the free (lattice) propagator. Using the so-called RI'
scheme we take 
\begin{equation}  
Z_q' \; = \; \frac{1}{12} \,
\left. tr \left( S^{-1}(p) \, \frac{\mbox{\boldmath{$\mathsf{1}$}}\, R(p) - i
\gamma_\mu a_\mu(p) }{R(p)^2 + \sum_\mu
a_\mu(p)^2} \right) \right|_{p^2=\mu^2} \;.
\end{equation}
Here $R$ and $a_\mu$ are the scalar and vector terms appearing in the
free CI Dirac operator, which in momentum space reads 
$D_{CI}(p) = i\gamma_\mu a_\mu(p) + R(p)$. 
Using Fourier transformation one can compute $R$ and $a_\mu$ from the 
definition and the parameters in $D_{CI}$ \cite{Ga01,GaLa03}.
They are normalized such that one finds
\begin{equation}
a_\mu(p) \; = \; i p_\mu + {\cal O}(a\,p)^2 \; \; \; \mbox{and}
\; \; \; R (p) \; = \; {\cal O}(a\,p)^2 \;.  
\label{eq_freedir}
\end{equation}
Landau gauge fixing $\partial_\mu A_\mu=0$ is implemented  as discussed 
in \cite{SuSc93,GiPaPa01} by iteratively minimizing a functional of the
link variables with stochastic overrelaxation \cite{MaOg90}. As is 
well known this type of gauge fixing still allows for Gribov copies. This
gives rise to an uncertainty which we address in Sect.\ \ref{sect_gribov}. 

It is easy to check that at tree level one finds
$\langle p\mid O_\Gamma\mid p\rangle_0 = \Gamma$.
Putting things together we obtain the final formula for $Z_\Gamma$ in the 
RI' scheme
\begin{equation}
Z_\Gamma^{\textrm{\scriptsize RI'}}\; = \;\left. 
\frac{12 \, Z_q'}{tr [ \, \Lambda_\Gamma(p) \,
\Gamma^{-1} \,]} \right|_{p^2=\mu^2} \; .
\label{finalform}
\end{equation}
For $\Gamma=\gamma_\nu, \gamma_\nu \gamma_5$ averaging over the index $\nu$
under the trace is implied.
The relation between different schemes is discussed in 
Sect.\ \ref{sect_conversions}.

\section{Technicalities}\label{sect_technicalities}

\subsection{Parameters}

The gauge configurations were generated in the quenched approximation
with the L\"uscher-Weisz action
\cite{LuWe85} at values of the gauge coupling $\beta=$ 7.90, 8.35 and 8.70
corresponding to lattice spacings of  
$a = 0.148\, \textrm{fm} = 0.750\, \textrm{GeV}^{-1}$, $a =
0.102\, \textrm{fm} = 0.517\, \textrm{GeV}^{-1}$ and $a = 0.078\, 
\textrm{fm} = 0.395\, \textrm{GeV}^{-1}$, respectively \cite{GaHoSc01}. 
The scale has been determined from the Sommer parameter $r_0$.

The final numbers we quote were computed on
lattices of size $16^3\times 32$. For studying possible volume and 
gauge fixing dependence we used additional ensembles on $8^3\times 24$ at
$\beta=7.90$. More details on the simulations can be found in \cite{GaGoHa04}.

For each lattice size and gauge coupling we determined the quark propagators
$S(x | p)$ on different (gauge fixed) gauge configurations and for
typically 16 different momentum sources. We used 11 different values 
for the quark masses, ranging from
$a\,m=$ 0.01 up to 0.20 covering a range of pion masses from 250 MeV 
up to 1-2 GeV (depending on $\beta$). The boundary 
conditions for the fermions were chosen
periodic in space and anti-periodic in the time direction. The momentum values
were taken to lie roughly along the diagonal of the Brillouin zone
ranging from the origin up to
$p=(5,5,6,10)\,2\,\pi/L$ with values of $a\,\mu$  up to 4.18.

For each mass and each operator $O_\Gamma$ we evaluated
$Z_\Gamma^{\textrm{\scriptsize RI'}}$ from (\ref{finalform}).  
On the small lattice we used 10 gauge configurations and on the larger
ones we had 5. We calculated the statistical error with the help of the
statistical bootstrap method.
% using 100 bootstrap samples.

\subsection{Extrapolation to the chiral limit}

\begin{figure}[tp]
  \begin{center}
    \includegraphics*[height=14.8cm]{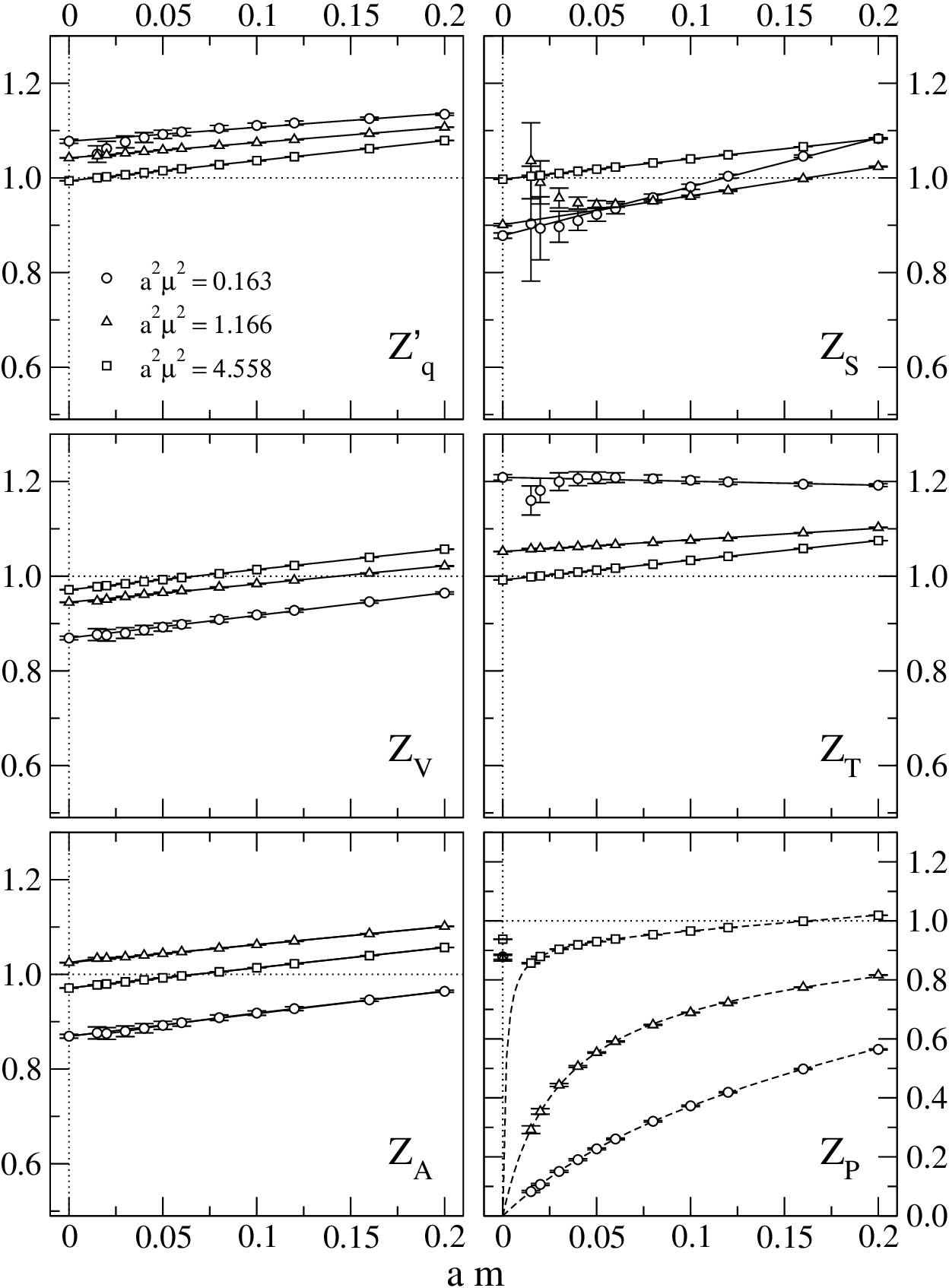}
  \end{center}
  \caption{\label{results_chiral_limit_plot} Results for
$\beta=7.90$ and volume $16^3\times 32$: The values for $Z_q'$ 
  (the fermion field renormalization constant), $Z_S$, $Z_V$,
$Z_T$, $Z_A$, $Z_P$  are plotted versus the quark mass at 
$a^2 \mu^2 = 0.163$ (circles), 1.166
(triangles), 4.558 (squares) together with extrapolations to the chiral limit.
For $Z_P$ the non-linear fit function (\ref{nonlinear_fit}) is 
included (dashed line).
The points at $m=0$ resulting from the corresponding subtracted 
extrapolation $Z_P^\textrm{\scriptsize{Sub}}(\mu^2)$ are also shown; 
the two values for the lower $a^2 \mu^2$ overlap.}
\end{figure}
As we want to obtain a mass-independent scheme 
we extrapolate all results for
$Z_\Gamma^{\textrm{\scriptsize RI'}}(\mu^2,m\,)$ linearly  
to the chiral limit $m\to 0$. In Fig.
\ref{results_chiral_limit_plot} this extrapolation is shown for three
representative values of $\mu^2$. Whereas the linear fit is appropriate for
most cases, we do find a substantially different behavior for the pseudoscalar
density. Similar observations have been discussed in 
\cite{GoHoOe99,Cu98,GiVl00,BeGiLu04}.

Since the pseudoscalar density couples to the Goldstone boson
channel there are ${\cal O}(1/m)$ contributions to $G_P$ in the chiral
limit.  We can incorporate this fact by expanding the inverse
renormalization  constant \cite{Cu98} in terms of the mass and adding the pole
term; in  \cite{Cu98} only a pole term and a constant term were used, but
one can also add terms ${\cal O}(m)$ and higher, 
which vanish in the chiral limit (cf.
\cite{BeGiLu04}). We use
\begin{equation}\label{nonlinear_fit}
  \frac{1}{Z_P(\mu^2, m)} \approx \frac{A(\mu^2)}{m} + B(\mu^2) + C(\mu^2)\,m.
\end{equation}
By subtracting the pole term first and then
performing the chiral limit we can define a ``subtracted''
renormalization constant
\begin{equation}
  \frac{1}{Z_P^\textrm{\scriptsize{Sub}}(\mu^2)} = 
  \lim_{m \to 0} \left( \frac{1}{Z_P(\mu^2)} - 
  \frac{A(\mu^2)}{m} \right) = \lim_{m \to 0} 
  \left( B(\mu^2) + C(\mu^2) m \right) = B(\mu^2)\;,
\end{equation}
such that the desired value is
\begin{equation}\label{ZPsubtracted}
  Z_P^\textrm{\scriptsize{Sub}}(\mu^2) = \frac{1}{B(\mu^2)}\;.
\end{equation}
The fit to $Z_P$ on the large lattices, as shown in Fig.
\ref{results_chiral_limit_plot}, is very satisfactory. For the small
lattice, which we use only to study the finite size dependence, the fit is
not as good for the smaller values of $\mu^2$ (cf.\ Subsection 
\ref{finite-size-effects}). 

For large $\mu$ the operator product expansion guarantees the suppression of
the pole contributions \cite{MaPiSa95}. This behavior is clearly exhibited 
in the plot for $Z_P$ when comparing the results 
for smaller $\mu^2$ to larger values of $\mu^2$. In terms of the 
expansion coefficients this means that $A(\mu^2)$
approaches 0 fast enough for higher $\mu^2$. The appropriate order 
of limits would be first $\mu\to\infty$ ($a\to 0$), then $m\to 0$. 
Here, due to the requirements of a
lattice calculation, we first perform $m\to 0$. 

The axial vector also couples to the Goldstone boson.
However, the coupling is proportional to the momentum transfer \cite{Pa79}, 
which vanishes for our kinematics.

For the further analysis we use the values in the chiral limit.

\subsection{Gauge fixing dependence}\label{sect_gribov}

\begin{figure}[t]
  \begin{center}
    \includegraphics*[width=12cm]{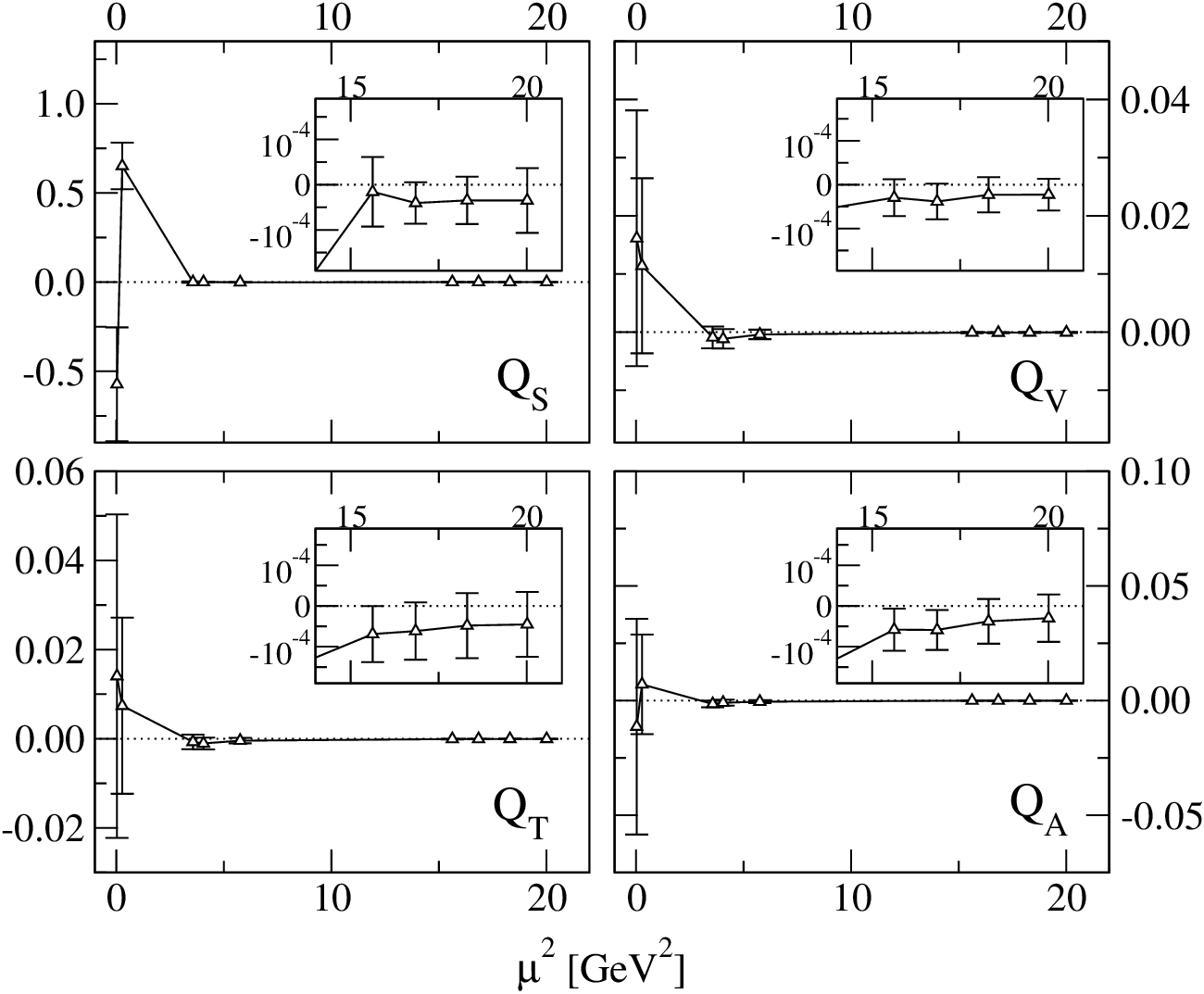}
  \end{center}
  \caption{\label{fig_gaugedependence} Values for the ratios
$Q_S$, $Q_V$, $Q_T$, $Q_A$, defined in Eq.\
(\protect{\ref{gribovratio}}),
vs.\ $\mu^2 [\textrm{GeV}^2]$ for the $8^3 \times 24$ lattice 
at $\beta = 7.90$.}
\end{figure}

Let us now analyze the effects due to gauge fixing
ambiguities. As discussed above we fix our gauge configurations to Landau
gauge using overrelaxation. If we apply a random gauge transformation 
to the original gauge configuration and then fix the gauge again, the 
procedure often leads to a different
(gauge-equivalent) configuration. When comparing the results for the
renormalization constants from the two gauge copies we find small
differences. This non-uniqueness is the Gribov uncertainty. We study the 
fluctuations induced by Gribov copies by comparing
results for $Z_\Gamma^{\textrm{\scriptsize RI'}}(\mu^2, m\,)$ 
derived with and without additional
random gauge transformations before applying the gauge fixing procedure. In
Fig.\ \ref{fig_gaugedependence} we plot the ratio 
\begin{equation}
  Q_\Gamma \; = \; \frac{Z_\Gamma^1 - Z_\Gamma^2}{Z_\Gamma^1 +
Z_\Gamma^2}\;,
\label{gribovratio}
\end{equation}
(where the superscript denotes the results from the two ``different'' 
configuration ensembles) as obtained for $8^3\times 24$ lattices at
$\beta=7.90$ with $Z_\Gamma^i$ taken in the chiral limit.  
For the values of $\mu^2$ of interest (around 4 $\textrm{GeV}^2$) 
we find  these
effects to be substantially smaller than 1\%. This result agrees with earlier
studies \cite{PaPeTa94,GiPeTa02a}.

\subsection{Finite size effects}\label{finite-size-effects}

\begin{figure}[tp]
  \begin{center}
    \includegraphics*[height=15cm]{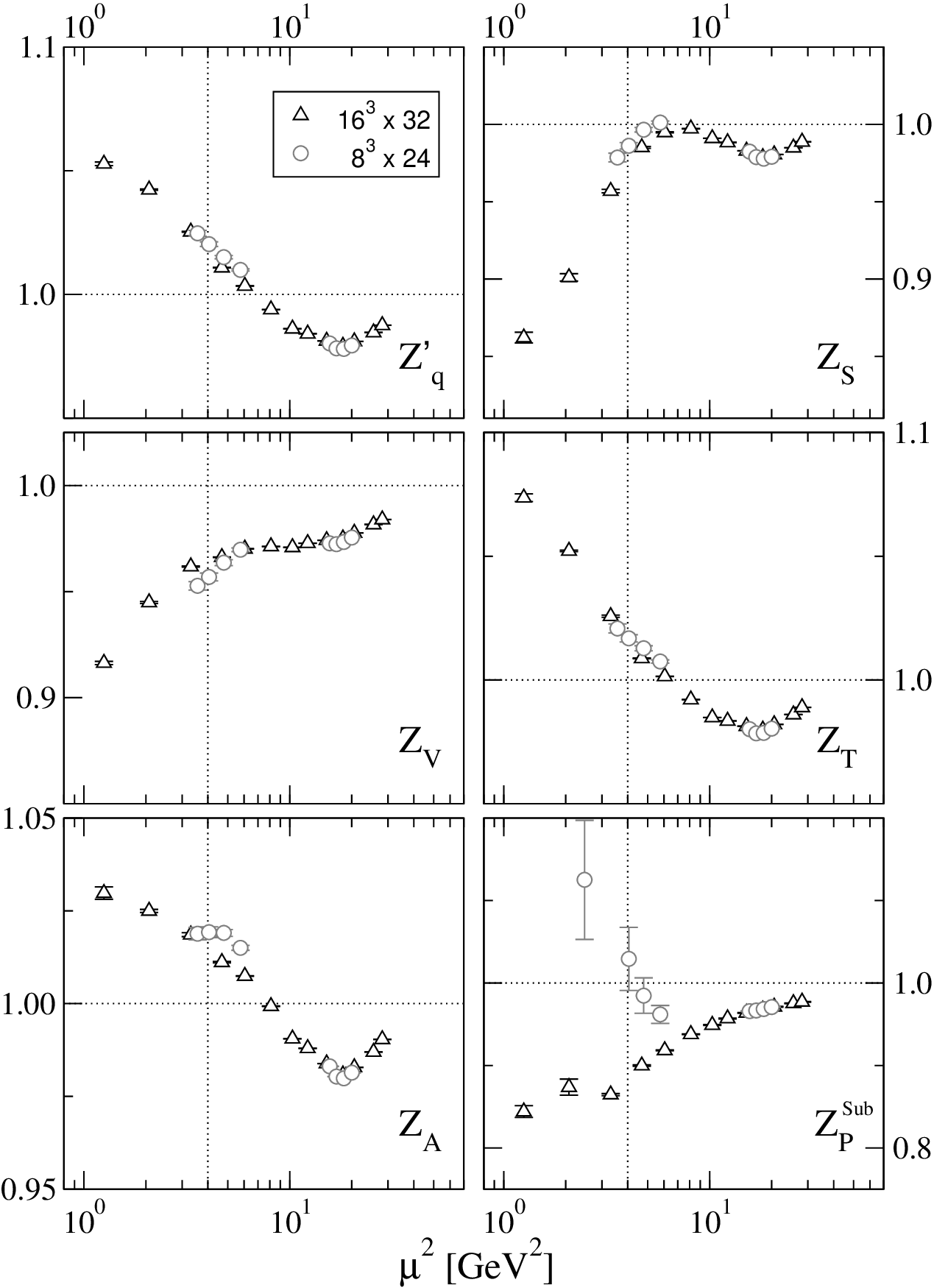}
  \end{center}
  \caption{\label{results_finite_volume_plot} Values for $Z_q'$, $Z_S$,
$Z_V$, $Z_T$, $Z_A$, $Z_P^\textrm{\scriptsize{Sub}}$ vs. 
$\mu^2$ at $\beta=7.90$. The triangles 
mark the values for the $16^3 \times 32$ lattice and the circles the
values for the $8^3 \times 24$ lattice. 
$Z_P^\textrm{\scriptsize{Sub}}$ is the value determined according 
to Eq.\ (\protect{\ref{ZPsubtracted}}).}
\end{figure}

We use two different lattice sizes in order to study the volume
dependence of the renormalization constants. In Fig.\
\ref{results_finite_volume_plot} we compare the values (extrapolated to
the chiral limit as discussed above) for the RI' scheme at $\beta=7.90$.
The deviations are generally small, increasing towards smaller momenta as
expected for finite size effects. However, 
above $\mu^2 \approx 4\,\textrm{GeV}^2$
the difference is generally less than 1\% except for 
$Z_P^\textrm{\scriptsize{Sub}}$. A possible explanation of this 
exceptional behavior might be that 
the coupling to the light pions enhances the finite size effects for
$Z_P^\textrm{\scriptsize{Sub}}$.
As mentioned in the discussion on the chiral limit, the extrapolation for $Z_P$
on the small lattices, in particular for small values of $\mu^2$, is not 
as good as on the larger lattices. This explains the larger error bars
for lower $\mu^2$ in this case.

We restrict ourselves to the
results from the largest lattice of size $16^3\times 32$ from now on.

\section{Results for RI'}\label{sect_results}

\begin{figure}[tp]
  \begin{center}
    \includegraphics*[height=15cm]{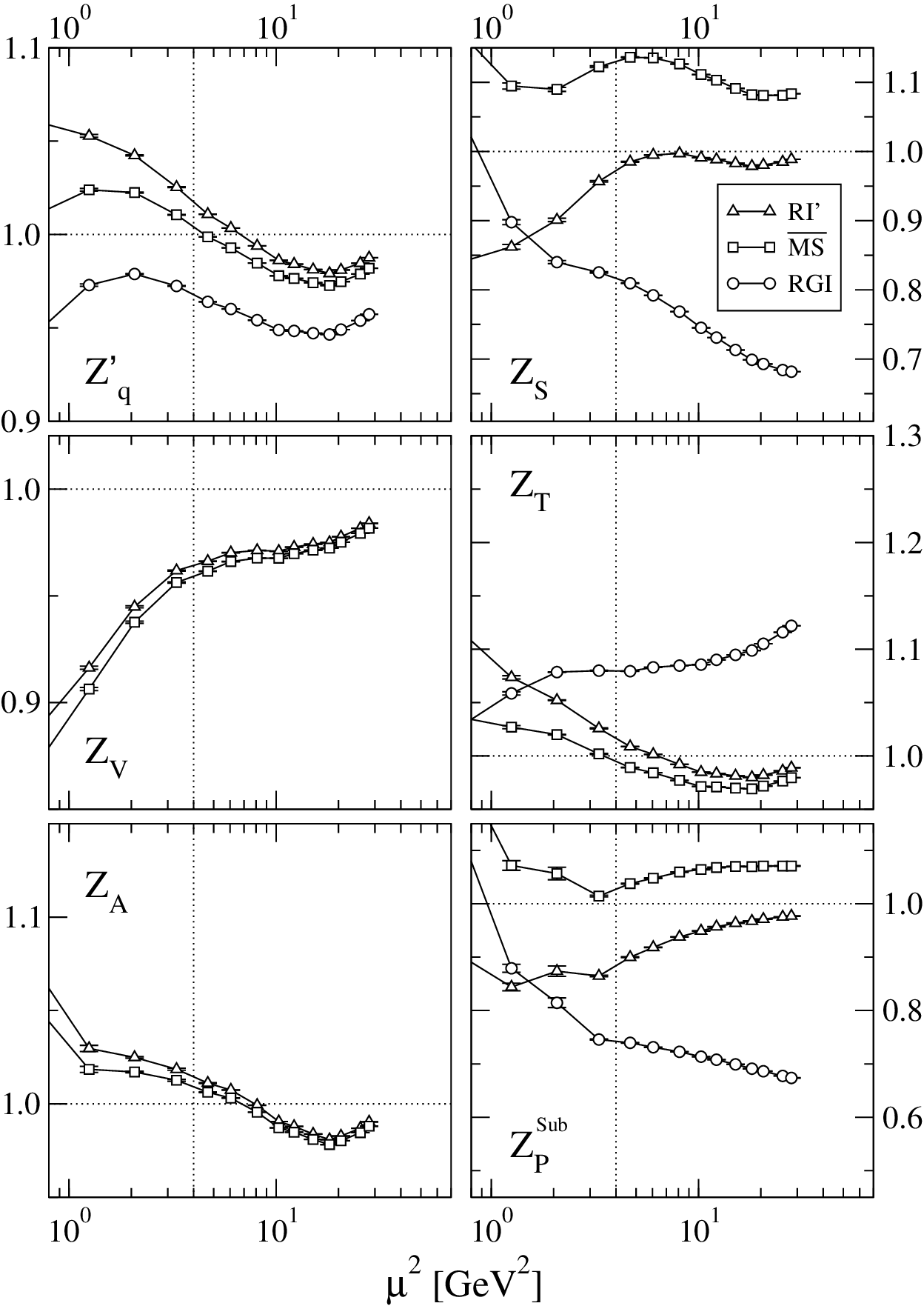}
  \end{center}
  \caption{\label{results_ZGamma_plot_large_790} 
Values for $Z_q'$, $Z_S$, $Z_V$,
$Z_T$, $Z_A$, $Z_P^\textrm{\scriptsize{Sub}}$ vs. 
$\mu^2 \,[\textrm{GeV}^2]$ for the $16^3 \times 32$ lattice at
$\beta = 7.90$. RI'\ values (triangles), \MSbar\ values (squares)
and RGI\ values (circles) are displayed. The dotted vertical line indicates
the value $\mu=2\,\textrm{GeV}$ where the 
\MSbar\ value has been obtained. Note that
for the vector and axial vector the \MSbar\ and the RGI values
coincide.}
\end{figure}

\begin{figure}[tp]
  \begin{center}
    \includegraphics*[height=15cm]{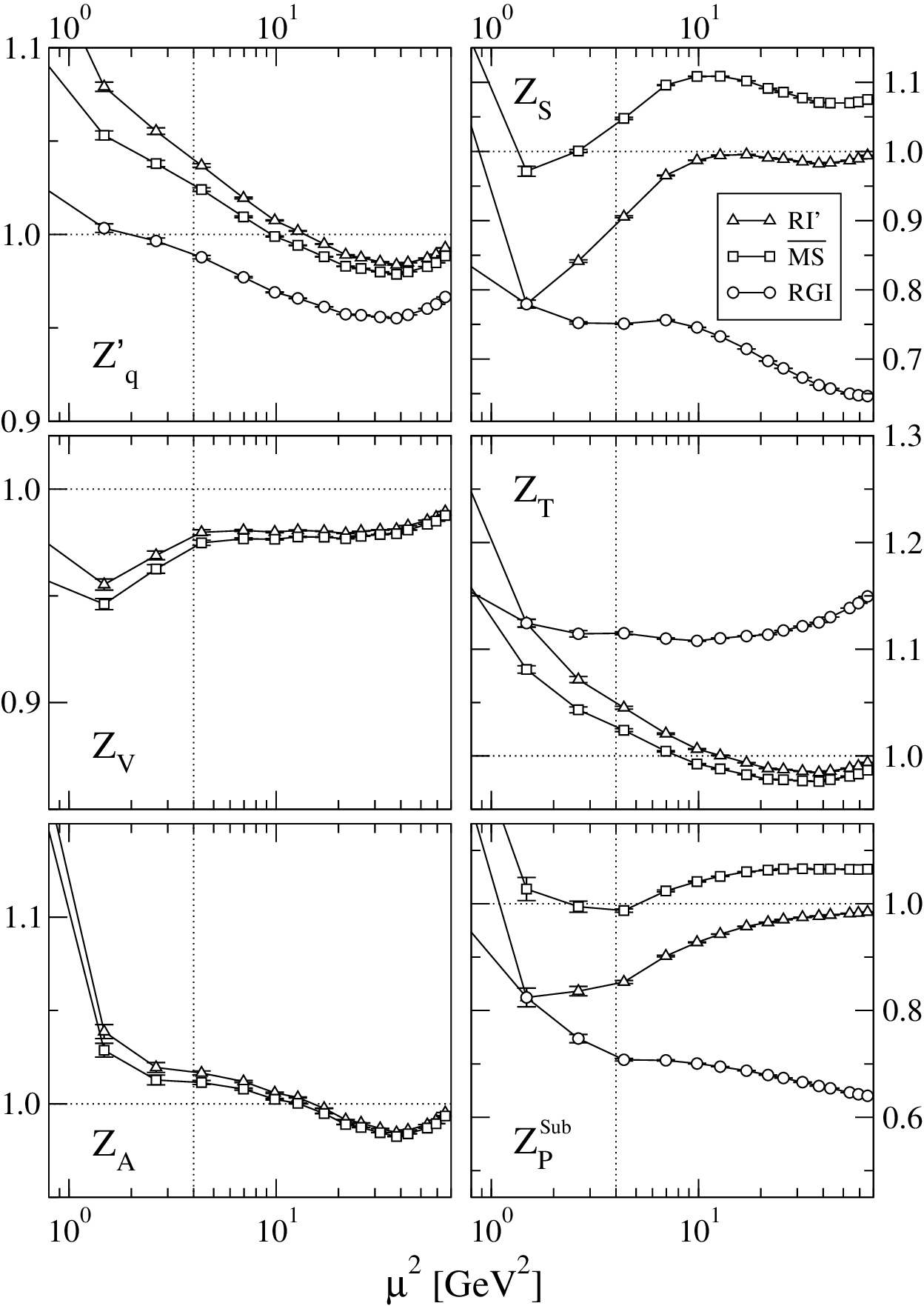}
  \end{center}
  \caption{\label{results_ZGamma_plot_large_835} 
Values for $Z_q'$, $Z_S$, $Z_V$,
$Z_T$, $Z_A$, $Z_P^\textrm{\scriptsize{Sub}}$ vs. 
$\mu^2 \,[\textrm{GeV}^2]$ for the $16^3 \times 32$ lattice at
$\beta = 8.35$. RI'\ values (triangles), \MSbar\ values (squares)
and RGI\ values (circles) are displayed. The dotted vertical line indicates
the value $\mu=2\,\textrm{GeV}$ where the \MSbar\ value has been obtained. 
Note that for the vector and axial vector the \MSbar\  and the RGI values
coincide.}
\end{figure}

\begin{figure}[tp]
  \begin{center}
    \includegraphics*[height=15cm]{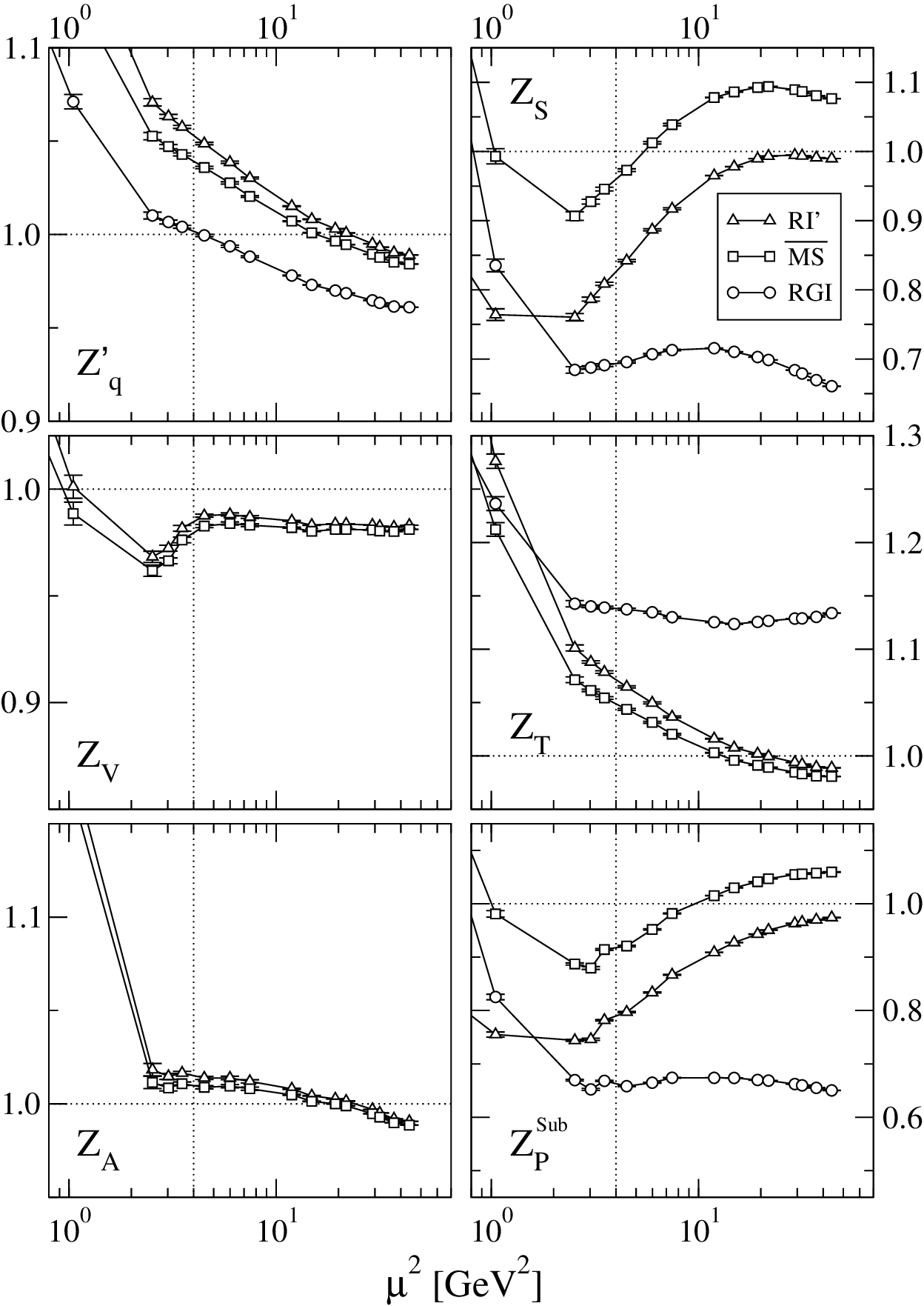}
  \end{center}
  \caption{\label{results_ZGamma_plot_large_870} 
Values for $Z_q'$, $Z_S$, $Z_V$, $Z_T$, $Z_A$, 
$Z_P^\textrm{\scriptsize{Sub}}$ vs. $\mu^2 \,[\textrm{GeV}^2]$ for the 
$16^3 \times 32$ lattice at
$\beta = 8.70$. RI'\ values (triangles), \MSbar\ values (squares)
and RGI\ values (circles) are displayed.  The dotted vertical line indicates
the value $\mu=2\,\textrm{GeV}$ where the \MSbar\ value has been obtained. 
Note that for the vector and axial vector the \MSbar\ and the RGI values
coincide.}
\end{figure}

In the appendix we collect our results for the chirally extrapolated
values of $Z_\Gamma^{\textrm{\scriptsize RI'}}$ in tabular
form. Figs.\ 
\ref{results_ZGamma_plot_large_790}-\ref{results_ZGamma_plot_large_870}
summarize these data. We discuss in this section only the RI' values
(triangles in Figs.\ 
\ref{results_ZGamma_plot_large_790}-\ref{results_ZGamma_plot_large_870}).

Vector and axial vector have finite renormalization factors and should 
be scale independent. Indeed the changes with $\mu^2$ (and also with 
the lattice size) are distinctly smaller than for the scalar or 
pseudoscalar renormalization factors. 
We also compared the variation of the renormalization factor for individual
components of $V_\mu$ and $A_\mu$ and the components of
$T_{\mu \nu}$. For 
$\mu^2\,>\,4 \,\textrm{GeV}^2$ the
deviation from the mean value is smaller than 1.5\,\% for all gauge couplings
studied, for $\mu^2\,>\,10 \,\textrm{GeV}^2$ it is less than 0.3\,\%. 
We therefore
plot and list the average over the results for these components according to
the remark following (\ref{finalform}).

Our currents are point-like and not the conserved currents. However, the
values for the renormalization constants for vector and axial vector are both
very close to 1, a behavior expected for conserved currents. 

\begin{figure}[tp]
\begin{center}
\includegraphics*[height=12cm]{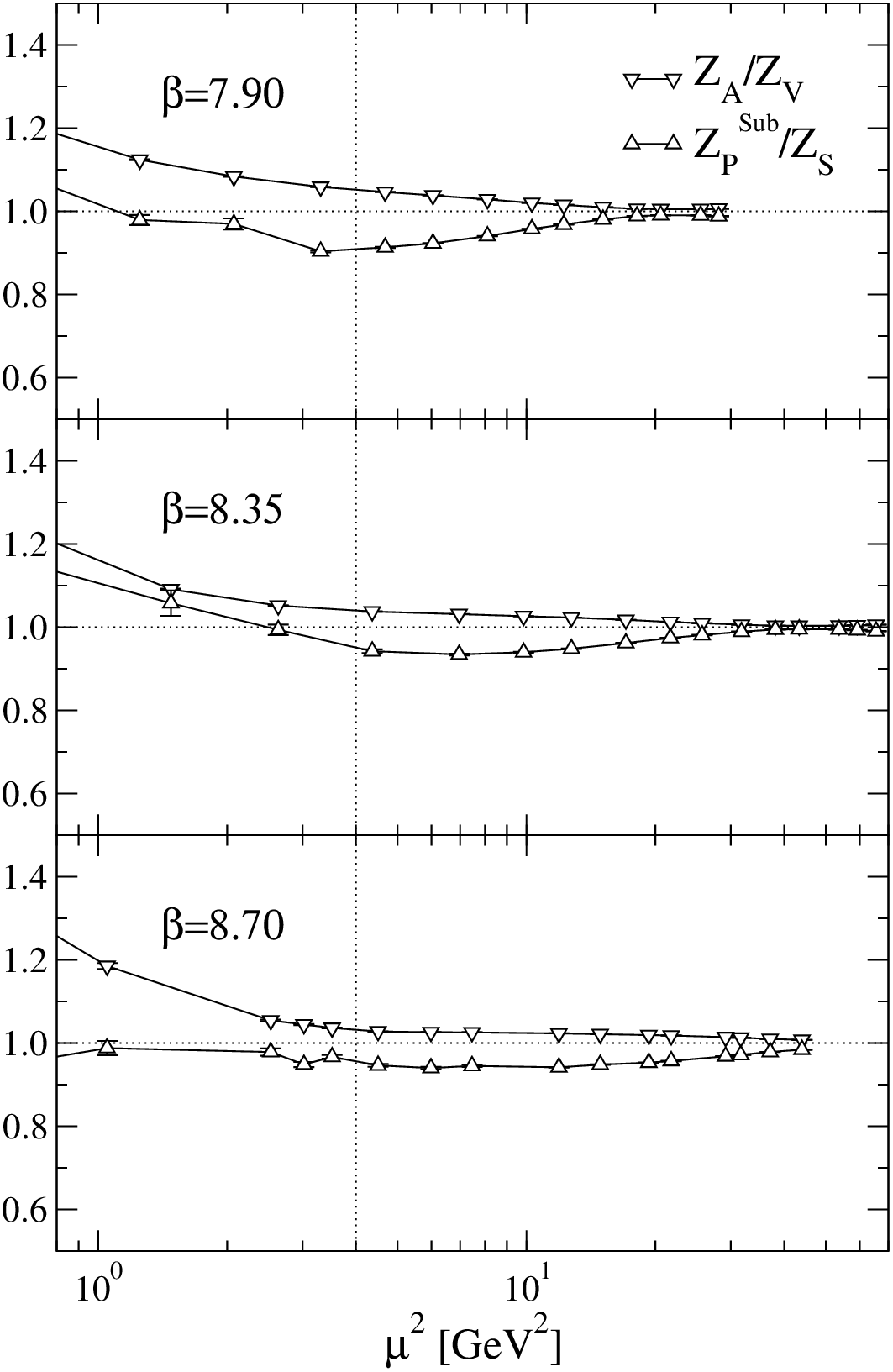}
\end{center}
\caption{\label{fig_ratio}Ratios $Z_A/Z_V$ and 
$Z_P^\textrm{\scriptsize{Sub}}/Z_S$ for the data 
in the chiral limit (lattice size $16^3\times 32$).
}
\end{figure}

The scalar and pseudoscalar densities are logarithmically divergent and have
scale dependent renormalization. Their ratio, however, should be scale
independent. Fig.\ \ref{fig_ratio} displays both ratios, $Z_A/Z_V$ and
$Z_P^\textrm{\scriptsize{Sub}}/Z_S$. Both ought to be 1 for chirally 
symmetric fermion actions.  
Indeed, the observed scale dependence of $Z_P^\textrm{\scriptsize{Sub}}$ and
$Z_S$ cancels in the ratio. Both ratios are surprisingly close to 1 with very 
little scale dependence, indicating the excellent approximation
to chiral symmetry of the considered CI fermion action.

Only for the two ensembles with smaller lattice spacing 
($\beta = 8.35, \beta = 8.7$) we have a reasonable chance to satisfy the
limits (\ref{mu_limits}). However, the relatively flat shape of the
scale independent $Z_A$ and $Z_V$ and the analysis of the $\mu$ 
dependence in the subsequent section indicate that we may still 
rely on the results  around
$\mu\approx 2\,\textrm{GeV}$. In fact, due to earlier results 
\cite{GaGoHa04} we expect
that the Dirac operator used here has improved scaling properties even for 
larger lattice spacings.

\section{Conversion to RGI and \MSbar}\label{sect_conversions}

\subsection{Different renormalization schemes}

The connection between different renormalization schemes is established 
using continuum perturbation theory.  
In the two schemes \MSbar\ and RI' the renormalization factors are related by
\begin{equation}
R_\Gamma (\mu^2) \; = \; 
\frac{Z_\Gamma^{\scriptsizeMSbar}(\mu^2)}{Z_\Gamma^{\textrm{\scriptsize RI'}}(\mu^2)}
     \;.
\end{equation}
Both, $Z_\Gamma^{\textrm{\scriptsize RI'}}(\mu^2)$ and 
$Z_\Gamma^{\scriptsizeMSbar}(\mu^2)$ 
are in general divergent in the continuum limit, the ratio stays finite, 
though. 

Note that the RI' scheme which we are employing differs from the RI scheme 
only by the definition of the quark field renormalization constant.  
The ratio of the latter in the two schemes 
\begin{equation}\label{Relation_RI-RIprime}
\frac{Z_q^{\textrm{\scriptsize RI}}}{Z_q'}
\; = \;\frac{Z_\Gamma^{\textrm{\scriptsize RI}}}
{Z_\Gamma^{\textrm{\scriptsize RI'}}}
\end{equation}
agrees with the conversion factor for the vector and axial vector 
renormalization constants from RI' to $\MSbar$ since there is no
further renormalization due to the protection by Ward identities.
In the Landau gauge and 3-loop order one has \cite{FrLu98,ChRe00}
\begin{equation}
\frac{Z_V^{\textrm{\scriptsize RI}}}
{Z_V^{\textrm{\scriptsize RI'}}}
=\frac{Z_A^{\textrm{\scriptsize RI}}}
{Z_A^{\textrm{\scriptsize RI'}}}
=1 - \frac{67}{6} \left( \frac{\alpha_s}{4 \pi} \right)^2
-  \left( 
\frac{52321}{72} -\frac{607 \,\zeta_3 }{4} \right)
\left( \frac{\alpha_s}{4 \pi} \right)^3
+ O(\alpha_s^4) \;.
\end{equation}
Here $\zeta_n$ is the Riemann zeta function evaluated at $n$.
This equation refers to the average over the different directional components
of the currents (under the trace in (\ref{finalform})). 
For the scalar and pseudoscalar one finds
\begin{eqnarray} \label{renormalization_definition_Rm}
\frac{Z_P^{\scriptsizeMSbar}}{Z_P^{\textrm{\scriptsize RI'}}} = 
\frac{Z_S^{\scriptsizeMSbar}}{Z_S^{\textrm{\scriptsize RI'}}}
& = & 1 + \frac{16}{3} \frac{\alpha_s}{4 \pi} + \left( \frac{4291}{18} 
- \frac{152 \zeta_3}{3} \right) \left( \frac{\alpha_s}{4 \pi} \right)^2 \\
 && +  \left( \frac{3890527}{324} - \frac{224993 \zeta_3}{54} + 
\frac{2960 \zeta_5}{9} \right) \left( \frac{\alpha_s}{4 \pi} \right)^3
+ O(\alpha_s^4) \nonumber \;.
\end{eqnarray}
In order to illustrate the orders of magnitude involved we give a numerical
value at the coupling $\alpha_s(\mu = 2\,\textrm{GeV}) = 0.203576$:
\begin{equation}
\frac{Z_S^{\scriptsizeMSbar}}{Z_S^{\textrm{\scriptsize RI'}}} 
\approx 1.16165 \; .
\end{equation}
The correction for the different definitions of the quark field
renormalization is rather small ($\approx$ 0.5\%), whereas 
even the $\alpha_s^3$ contribution to the scheme conversion
is large ($\approx$ 2.9\,\%) in comparison. 

Concerning the tensors, the conversion factor from RI' to \MSbar\ 
has been computed in \cite{Gr03}. 
In Landau gauge one finds
\begin{eqnarray}
 \frac{Z_T^{\scriptsizeMSbar}}{Z_T^{\textrm{\scriptsize RI'}}}
 & = &  1 + \left( -\frac{3847}{54} + 
\frac{184 \zeta_3}{9} \right) \left( \frac{\alpha_s}{4 \pi} \right)^2 + \\
 & &    \left( -\frac{9858659}{2916} + \frac{678473 \zeta_3}{486} + \frac{1072 
\zeta_4}{81} - \frac{10040 \zeta_5}{27} \right) 
\left( \frac{\alpha_s}{4 \pi} \right)^3 + 
\mathcal{O}(\alpha_s^3). \nonumber
\label{renormalization_definition_RT}
\end{eqnarray}

We compute the coupling $\alpha_s$ from the 3-loop expression in the 
\MSbar\ scheme
\begin{eqnarray}
\frac{\alpha_s (q^2)}{4 \pi} 
& = &\frac{1}{\beta_0 \log(q^2)} - 
\frac{\beta_1}{\beta_0^3} \frac{\log \log (q^2)}{\log^2(q^2)} \\ \nonumber
 && + \frac{1}{\beta_0^5 \log^3(q^2)} \left( \beta_1^2 \log^2 \log(q^2) 
- \beta_1^2 \log \log (q^2) + 
\beta_2^{\scriptsizeMSbar} \beta_0 - \beta_1^2 \right) \;,
\end{eqnarray}
where we have introduced the shorthand notation
$q^2 = \mu^2/(\Lambda_\textrm{\scriptsize QCD}^{\scriptsizeMSbar})^2$.
For $\Lambda_\textrm{\scriptsize QCD}^{\scriptsizeMSbar}$ 
in the quenched approximation
we use the value 
$\Lambda_\textrm{\scriptsize QCD}^{\scriptsizeMSbar} = 0.238 \pm 0.019\, 
\mbox{GeV}$ calculated in \cite{CaLuSo99}.

In Figs.\ 
\ref{results_ZGamma_plot_large_790}-\ref{results_ZGamma_plot_large_870}
we plot our results converted to the 
\MSbar\ scheme. Except for the vector currents we expect dependence on 
both, $\mu^2$ and lattice spacing.
 
In Tables 
\ref{results_table_ZRI_large_790}-\ref{results_table_ZRI_large_870}
the \MSbar\ values are given at $\mu = 2\,\textrm{GeV}$ determined by linear
interpolation between the closest $\mu$\ values of the  data.

\subsection{RGI values}

The renormalization factors depend on the renormalization scale $\mu$
(except for the vector and axial currents). This dependence is governed
by the anomalous dimension $\gamma$ of the operator:
\begin{equation}
\gamma (\alpha_s) = - \mu \frac{\mathrm d}{\mathrm d \mu} \ln Z (\mu^2)
= \sum_{i=0}^\infty \gamma_i 
   \left( \frac{\alpha_s (\mu^2)}{4 \pi} \right)^{i+1} \;.
\end{equation}
Integrating this differential equation
we can define a scale-independent quantity 
$Z^{\textrm{\scriptsize RGI}}$ (RGI = Renormalization Group Invariant)
by \cite{GiGiRa98}
\begin{equation}
Z^{\textrm{\scriptsize RGI}} = Z(\mu^2) 
\left( 2 \beta_0 \frac{\alpha_s(\mu^2)}{4 \pi} \right)^{-\gamma_0/(2 \beta_0)}
  \exp \left \{ \frac{1}{2} \int_0^{\alpha_s(\mu^2)} \! \mathrm d \alpha
    \left( \frac{\gamma (\alpha)}{\beta (\alpha)} + 
           \frac{\gamma_0}{\beta_0 \alpha} \right)
   \right \} \;,
\end{equation}
where the $\beta$ function is given by
\begin{equation}
\beta (\alpha_s)= 
 \frac{\mu}{2} \frac{\mathrm d}{\mathrm d \mu} \alpha_s (\mu^2)
 = - 4 \pi \sum_{i=0}^\infty \beta_i 
    \left( \frac{\alpha_s(\mu^2)}{4 \pi} \right)^{i+2} \;.
\end{equation}
Since we know the renormalization group functions $\beta$ and $\gamma$ 
only to a certain order in perturbation theory we cannot expect exact
scale invariance when we apply the above definitions to our data. 
However, we can check how well the expectation of
$\mu$-independence is met \cite{GoHoOe99}.

In Figs.\ 
\ref{results_ZGamma_plot_large_790}-\ref{results_ZGamma_plot_large_870} we
also show the RGI\ values as obtained by converting at each 
value of $\mu^2$ (indicated
by circles). For perfect 3-loop scaling these values should be 
constant; they are not,
but they have indeed a plateau-like behavior for 
$ 2 \,\textrm{GeV} <\mu < 4 \,\textrm{GeV}$ with  less than
5\% variation (10\% for $Z_S^{\textrm{\scriptsize RGI}}$).

In Tables 
\ref{results_table_ZRI_large_790}-\ref{results_table_ZRI_large_870}
we quote the RGI\ values at $a \mu =1.08$ 
(the actual values of $a \mu$ closest to 1). 

\section{Summary and conclusion}\label{sect_conclusion}

We have determined the renormalization constants of bilinear 
quark operators for the chirally improved lattice Dirac operator by 
non-perturbative methods. These allow us to relate certain lattice 
observables to those in continuum renormalization schemes.
We observe reasonable agreement with 3-loop renormalization group behavior, 
which improves as the lattice spacing becomes smaller.
Finite size effects and problems from Gribov copies were found to be 
negligible. The Goldstone boson pole in the renormalization factor $Z_P$ of
the pseudoscalar density could be subtracted leaving behind a well-behaved
$Z_P^\textrm{\scriptsize{Sub}}$. The
renormalization constants for the vector and the axial vector currents are
close to 1. The ratios $Z_A/Z_V\approx 1.03$ and 
$Z_P^\textrm{\scriptsize{Sub}}/Z_S\approx 0.95$ 
for the smallest lattice 
spacing indicate a very good approximation of chiral symmetry.
\\
\\
\\ 
{\bf Acknowledgments:}
The quark propagators have been determined on the Hitachi SR8000 of the
Leibniz Rechenzentrum Munich. We wish to thank Vladimir Braun, Christian
Hoelbling, Stefan Schaefer, 
and Andreas Sch\"afer for interesting discussions and support.
Support by Fonds zur F\"orderung der Wissenschaftlichen Forschung in
\"Osterreich, project FWF P16310-N08, by \"Osterreichische Akademie der
Wissenschaften (APART 654), DFG (Forschergruppe 
``Gitter-Hadronen-Ph\"anomenologie''), and BMBF is gratefully acknowledged.

\clearpage

\begin{appendix}
\section{Tables of results}

%%% table RGI 16x32, beta=7.90 %%%
\begin{table}[htf]
\begin{center}
\begin{tabular}{r|l|l|l|l|l}

  $\mu^2 \,[\textrm{GeV}^2]$ &    $Z_S$ & $Z_V$ & $Z_T$ & $Z_A$ & $Z_P^\textrm{\scriptsize{Sub}}$ \\ \hline

%cat ../../tabs/Z.16x32_b7.90.RI.tex
  0.017 &       0.7(1) &        0.65(8) &       1.0(1) &        1.1(1) &        0.98(2) \\ 
  0.291 &       0.878(6) &      0.869(4) &      1.209(6) &      1.180(4) &      0.878(9) \\ 
  0.702 &       0.84(1) &       0.887(2) &      1.118(3) &      1.071(5) &      0.90(2) \\ 
  1.251 &       0.862(3) &      0.9163(8) &     1.074(2) &      1.030(2) &      0.844(7) \\ 
  2.073 &       0.901(2) &      0.9449(5) &     1.0522(3) &     1.0249(4) &     0.87(1) \\ 
  3.307 &       0.957(1) &      0.9618(3) &     1.0258(4) &     1.0186(5) &     0.865(1) \\ 
  4.677 &       0.9851(6) &     0.9662(1) &     1.0087(2) &     1.0112(2) &     0.8999(8) \\ 
  6.048 &       0.9948(3) &     0.9702(1) &     1.00145(6) &    1.00744(7) &    0.9183(5) \\ 
  8.104 &       0.9972(1) &     0.97129(8) &    0.99205(6) &    0.99930(6) &    0.9379(4) \\ 
  10.297 &      0.99097(7) &    0.97088(3) &    0.98483(2) &    0.99055(2) &    0.9491(4) \\ 
  12.216 &      0.98833(5) &    0.97282(3) &    0.98345(2) &    0.98797(2) &    0.9570(5) \\ 
  15.095 &      0.98289(3) &    0.97420(3) &    0.98132(2) &    0.98385(1) &    0.9638(4) \\ 
  18.110 &      0.97871(2) &    0.97502(2) &    0.97990(1) &    0.98088(1) &    0.9676(4) \\ 
  20.578 &      0.98051(2) &    0.97767(2) &    0.98207(2) &    0.98282(1) &    0.9712(4) \\ 
  25.512 &      0.98491(1) &    0.98165(2) &    0.98603(1) &    0.98698(1) &    0.9754(4) \\ 
  28.116 &      0.98870(1) &    0.98398(2) &    0.98893(2) &    0.99038(2) &    0.9772(4) \\ 
  \hline
%cat ../../tabs/Z.16x32_b7.90.RGI-Goeckeler-order3.tex
  RGI & 0.840(2) &      0.9376(5) &     1.0785(3) &     1.0171(4) &     0.815(9) \\ 
  \hline
%cat ../../tabs/Z.16x32_b7.90.MS-value-order3.tex
  \MSbar &      1.1309(9) &     0.9586(2) &     0.9944(3) &     1.0087(4) &     1.0281(5) \\ 

\end{tabular}
\end{center}
\caption{\label{results_table_ZRI_large_790} Values for the renormalization
constants in the RI' scheme for the $16^3 \times 32$ lattice at
$\beta = 7.90$ as resulting from the extrapolation to $m=0$. The RGI value is
taken at $\mu^2 = 2.073\, \textrm{GeV}^2$, i.e. $a \mu = 1.07992$.
The $\MSbar\,(2\, \textrm{GeV})$ value is computed by interpolation.}
\end{table}

%%% table RGI 16x32, beta=8.35 %%%

\newpage
\begin{table}[htf]
\begin{center}
\begin{tabular}{r|l|l|l|l|l}

  $\mu^2 \,[\textrm{GeV}^2]$ &    $Z_S$ & $Z_V$ & $Z_T$ & $Z_A$ & $Z_P^\textrm{\scriptsize{Sub}}$ \\ \hline

%cat ../../tabs/Z.16x32_b8.35.RI.tex
  0.036 &       1.7(1) &        0.99(1) &       1.61(3) &       1.95(5) &       1.9(3) \\ 
  0.613 &       0.86(2) &       0.982(4) &      1.301(8) &      1.22(1) &       1.00(6) \\ 
  1.479 &       0.779(6) &      0.955(3) &      1.124(4) &      1.039(4) &      0.82(2) \\ 
  2.633 &       0.842(2) &      0.969(2) &      1.072(3) &      1.020(3) &      0.836(9) \\ 
  4.365 &       0.906(1) &      0.980(1) &      1.045(1) &      1.017(1) &      0.853(3) \\ 
  6.962 &       0.9653(6) &     0.9806(4) &     1.0209(6) &     1.0120(6) &     0.902(2) \\ 
  9.848 &       0.9873(3) &     0.9799(3) &     1.0065(4) &     1.0060(4) &     0.9277(8) \\ 
  12.733 &      0.9945(2) &     0.9807(3) &     1.0004(3) &     1.0034(3) &     0.9431(3) \\ 
  17.062 &      0.9955(2) &     0.9802(1) &     0.9936(1) &     0.9976(1) &     0.9576(4) \\ 
  21.679 &      0.99084(6) &    0.97923(8) &    0.98829(7) &    0.99153(7) &    0.9653(5) \\ 
  25.719 &      0.98908(5) &    0.98027(6) &    0.98728(5) &    0.98978(5) &    0.9704(5) \\ 
  31.780 &      0.98535(3) &    0.98090(4) &    0.98563(3) &    0.98683(3) &    0.9747(5) \\ 
  38.128 &      0.98235(2) &    0.98116(3) &    0.98446(2) &    0.98464(2) &    0.9769(5) \\ 
  43.323 &      0.98373(2) &    0.98284(3) &    0.98589(2) &    0.98600(2) &    0.9791(5) \\ 
  53.712 &      0.98695(3) &    0.98538(3) &    0.98854(3) &    0.98892(3) &    0.9818(5) \\ 
  59.194 &      0.98961(4) &    0.98687(3) &    0.99047(3) &    0.99125(3) &    0.9828(5) \\ 
  65.543 &      0.99425(4) &    0.98943(3) &    0.99383(3) &    0.99532(4) &    0.9847(5) \\ 
  \hline
%cat ../../tabs/Z.16x32_b8.35.RGI-Goeckeler-order3.tex
  RGI & 0.751(1) &      0.975(1) &      1.115(1) &      1.011(1) &      0.708(2) \\ 
  \hline
%cat ../../tabs/Z.16x32_b8.35.MS-value-order3.tex
  \MSbar &      1.039(1) &      0.973(1) &      1.028(2) &      1.012(1) &      0.987(4) \\ 

\end{tabular}
\end{center}
\caption{\label{results_table_ZRI_large_835} Values for the renormalization
constants in the RI' scheme for the $16^3 \times 32$ lattice at
$\beta = 8.35$ as resulting from the extrapolation to $m=0$. The RGI value is
taken at $\mu^2 = 4.365\, \textrm{GeV}^2$, i.e. $a \mu = 1.07992$.
The $\MSbar \,(2\, \textrm{GeV})$ value is computed by interpolation.}
\end{table}

\newpage
%%% table RGI 16x32, beta=8.70 %%%

\begin{table}[htf]
\begin{center}
\begin{tabular}{r|l|l|l|l|l}

  $\mu^2 \,[\textrm{GeV}^2]$ &    $Z_S$ & $Z_V$ & $Z_T$ & $Z_A$ & $Z_P^\textrm{\scriptsize{Sub}}$ \\ \hline

%cat ../../tabs/Z.16x32_b8.70.RI.tex
  0.062 &       1.73(9) &       1.08(1) &       1.84(3) &       2.16(5) &       0.64(4) \\ 
  0.555 &       0.89(2) &       1.078(8) &      1.52(1) &       1.44(1) &       0.838(7) \\ 
  1.049 &       0.764(8) &      1.001(5) &      1.276(7) &      1.180(7) &      0.755(5) \\ 
  2.529 &       0.760(5) &      0.968(3) &      1.101(3) &      1.018(3) &      0.744(2) \\ 
  3.023 &       0.787(3) &      0.972(1) &      1.088(1) &      1.015(2) &      0.746(2) \\ 
  3.516 &       0.809(2) &      0.982(1) &      1.079(1) &      1.0165(9) &     0.782(1) \\ 
  4.503 &       0.842(2) &      0.9876(7) &     1.065(1) &      1.0139(8) &     0.797(1) \\ 
  5.984 &       0.887(1) &      0.9881(7) &     1.0496(9) &     1.0139(9) &     0.834(1) \\ 
  7.464 &       0.917(2) &      0.9869(7) &     1.0366(8) &     1.0121(9) &     0.8669(8) \\ 
  11.905 &      0.9652(3) &     0.9851(2) &     1.0161(3) &     1.0081(3) &     0.9090(2) \\ 
  14.866 &      0.9780(2) &     0.9831(1) &     1.0078(2) &     1.0043(2) &     0.9274(2) \\ 
  19.308 &      0.9896(2) &     0.98369(8) &    1.0020(1) &     1.0027(1) &     0.9431(4) \\ 
  21.775 &      0.9933(2) &     0.98363(8) &    0.9996(1) &     1.0015(1) &     0.9506(3) \\ 
  29.177 &      0.99479(8) &    0.98312(3) &    0.99397(5) &    0.99705(7) &    0.9632(4) \\ 
  31.645 &      0.99373(8) &    0.98255(3) &    0.99217(5) &    0.99514(6) &    0.9655(4) \\ 
  37.073 &      0.99102(5) &    0.98225(2) &    0.98963(3) &    0.99202(4) &    0.9699(4) \\ 
  43.982 &      0.98976(3) &    0.98310(2) &    0.98881(2) &    0.99064(2) &    0.9741(4) \\ 
  \hline
%cat ../../tabs/Z.16x32_b8.70.RGI-Goeckeler-order3.tex
  RGI & 0.713(1) &      0.9831(7) &     1.1300(9) &     1.0081(8) &     0.6739(6) \\ 
  \hline 
%cat ../../tabs/Z.16x32_b8.70.MS-value-order3.tex
  \MSbar &      0.959(2) &      0.979(1) &      1.049(1) &      1.0095(7) &     0.915(1) \\ 
 
\end{tabular}
\caption{\label{results_table_ZRI_large_870} Values for the renormalization
constants  in the RI' scheme for the $16^3 \times 32$ lattice at
$\beta = 8.70$ as resulting from the extrapolation to $m=0$. The RGI value is
taken at $\mu^2 = 7.464\, \textrm{GeV}^2$, i.e. $a \mu = 1.07992$.
The $\MSbar\,(2\, \textrm{GeV})$ value is computed by interpolation.}
\end{center}
\end{table}

\end{appendix}

\newpage

%%%%%%%%%%%%%%%%%%%%%%%%%%%%%%%%%%%%%%%%%%%%%%%%%%%%%%%%%%%%%%%%%%%%%%%%%%%
% BIBLIOGRAPHY
%%%%%%%%%%%%%%%%%%%%%%%%%%%%%%%%%%%%%%%%%%%%%%%%%%%%%%%%%%%%%%%%%%%%%%%%%%%
%\bibliographystyle{/home/cbl/refs/bibtex/npb}
%\bibliography{/home/cbl/refs/Lgt+2000,/home/cbl/refs/Lgt-1999,/home/cbl/refs/Lgt-1989}

\end{document}